\documentclass[12pt]{article} \usepackage{subfigure} \usepackage{cite}
\usepackage{latexsym} \usepackage{amssymb} \textwidth 16cm
\usepackage{amsmath}
\usepackage{pictex}
\usepackage{graphics}
\newcommand{\swn}{small-world network}
\newcommand{\eq}[1]{Eq.~(\ref{#1})}
\begin{document}
\title{Small-World Networks: Links with long-tailed distributions.}
\author{ S. Jespersen$^{a,b}$\\
A. Blumen$^{b}$\\
$^a$Institute of Physics and  Astronomy\\
University of Aarhus, DK-8000 {\AA}rhus C\\
$^b$Theoretische Polymerphysik\\ Universit\"at Freiburg D-79104
Freiburg i.Br., Germany }

\date{\today}  
\maketitle
\begin{abstract}
Small-world networks (SWN), obtained by randomly adding to a regular
structure additional links (AL), are of current interest. In this article
we explore (based on physical models) a new variant of SWN, in which the
probability of realizing an AL depends on the chemical distance
between the connected sites. We assume a power-law probability
distribution and study random walkers on the network,
focussing especially on their probability of being at the origin. We
connect the results to L\'evy Flights, which follow from a mean field
variant of our model.
\end{abstract}
\parbox{1\linewidth}{PACS numbers: 05.40Fb, 05.60.-k, 71.55.Jv.}
\section{Introduction}\label{intro}
Recently a lot of interest has centered on the so-called {\swn
}s (SWN)\cite{watts,newman4} where an underlying regular lattice is
supplemented with additional links  (bonds), a fact which drastically
reduces the minimal distances (the so-called chemical distances in the
fractal literature) between pairs of points on the
lattice\cite{watts,barrat,moukarzel2,newman2,newman3,kulkarni,barthelemy}. This
question is of utmost importance  for general network structures, say
Internet-links\cite{albert}, and for the spreading of
diseases\cite{newman1,moukarzel2,pandit,moore,kleczkowski}, topics
which depend on the minimal distances. On the other hand other
questions are envisageable over  such structures, for example random
transport\cite{sj1}; this requires solving diffusion-type problems,
which are mathematically described by the Laplacian on the
structure\cite{monasson} and the corresponding eigenvalues and
eigenvectors\cite{parbati}. Examples of such problems are anomalous
transport of charges and of 
excitations over networks\cite{sokolov}.  Most recent SWN-studies
center on a one-dimensional chain supplemented with additional links (AL),
which connect sites that are arbitrarily far from each other on the
underlying lattice. While this being the simplest SWN envisageable,
there are situations in which links between distant sites occur
naturally; however, their lengths are then not necessarily uniformly
distributed: Considering a 
polymer chain in solution, monomers which are far apart along the
backbone can be quite close to each other in real space, so that for
instance energy transfer over the structure may take crosscuts along
sites near to each other in space\cite{sokolov}. Now the probability
of having such close monomer pairs is related to the return to the origin of
random walks, possibly under self-avoiding constraints. In this case
the probability $p(l)$ that two sites far 
apart along the backbone get close together in space, goes
approximately as an inverse power-law of their mutual chemical
distance $l$\cite{doi}:
\begin{equation}
\label{powerlaw}
p(l)=\frac{a}{l^{\alpha}}
\end{equation}
In this work, we will focus on networks constructed as in the SWN
case, while however letting the additional bonds be distributed
according to Eq. (\ref{powerlaw}). We call these structures
generalized small-world networks (GSWN). Clearly, the original SWN is
recovered from the GSWN by letting $\alpha\rightarrow 0$. On the other
hand, in GSWN with $\alpha \gg 1$ practically only sites which are
already very close on the chain get to be connected; such GSWN have 
(apart from disorder) properties similar to the underlying regular
lattice. Most interesting are GSWN with $0\leq\alpha\leq 3$, which
show a wealth of features, because of the long-range character of the
additional links.  

In this paper we will study random walks over GSWN, and
especially the probability $P_0(t)$ of the walker being (still or
again) at the origin; as 
discussed in previous works\cite{sj1,sj2}, this quantity reflects many
of the properties of the density of eigenvalues of the underlying
structure, and is easily obtainable by very effective, easy to program
numerical procedures.

Our paper is structured as follows: In the next section
(Sec.~\ref{def}) we discuss the construction of GSWN in more
detail. In Sec.~\ref{rel} we study the behavior of random walkers on
GSWN. We find that for $\alpha$ well below $2$ we have a behavior
qualitatively similar to that of walkers over the SWN. However,  for
$\alpha$ larger than $2$ we move towards another regime, quite
reminiscent of random walks on regular lattices. The transition
appears to happen around $\alpha\approx 2$, which prompts us to
consider transient  versus recurrent walks in Sec.~\ref{discuss}. In
this Section we determine analytically $P_0(t)$ for a mean field
variant of the GSWN model, which we then compare with the numerical
findings $P_0(t)$ on GSWN. It turns out that the mean-field approach
is related to L\'evy Flights and L\'evy Walks. Finally we close our
paper by summarizing our conclusions in Sec.\ref{conclusion}.   

\section{Construction of GSWN}\label{def}
\begin{figure}[h]
\unitlength=1cm
\begin{center}
\begin{picture}(6,5.8)
\put(-2.2,6.6){ \includegraphics{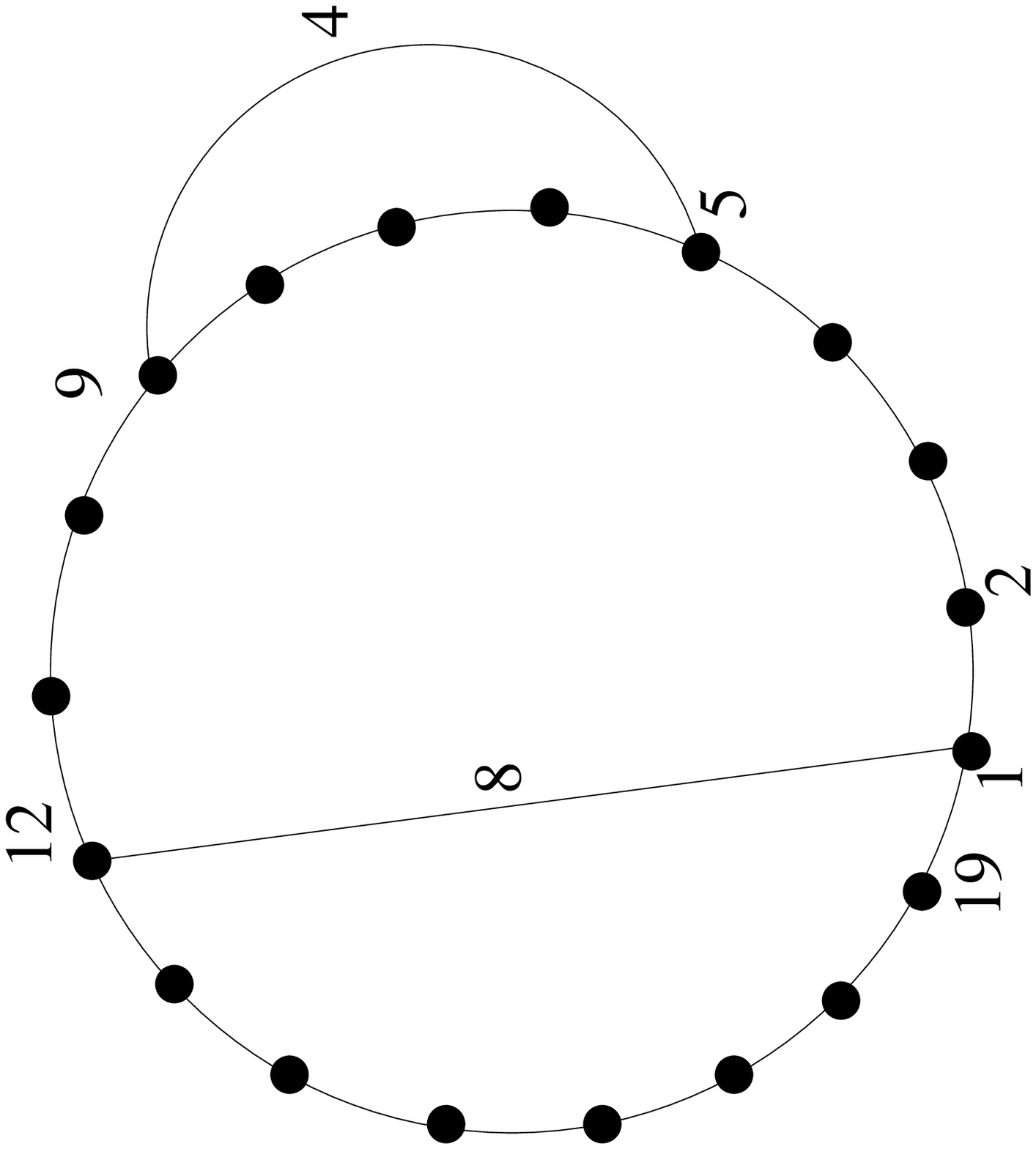}}
\end{picture}
\end{center}
\caption{Illustration of the \swn. In this example two additional
  links are added, with the corresponding distances given on the
  figure.}
\label{ring}
\end{figure}

The construction of GSWN follows at first the SWN procedure closely:
We start from a ring of $N$ sites  (i.e. a closed, regular $1$d
lattice). Then we consider each site consecutively, and let it sprout
with probability $q$ an additional bond, which connects it to another
site, see Fig.~\ref{ring}. We now let $p(l)$ in \eq{powerlaw} be the
conditional probability that this bond gets attached to a site at the
(minimal) chemical distance $l$ from the sprouting site, measured
along the ring. Here, given our periodic
boundary conditions, the chemical distance $l$ lies between $1$ and
int$(N/2)$, where int$(x)$ denotes the largest integer $X$ such that
$X\leq x$ (see Fig. \ref{ring}). Note that through $p(l)$ our model
differs from the standard SWN, where no $l$ dependence is accounted
for. In \eq{powerlaw}, the constant $a$ normalizes the distribution so
that 
\begin{equation}
\label{norm}
2\sum_{l=1}^{int(N/2)} al^{-\alpha}=1
\end{equation}
Now the exponent $\alpha$ in \eq{powerlaw} is a parameter of the model
and will be varied in the following. For a finite system we can choose
it freely, so that $\alpha\geq 0$. In fact the choice $\alpha=0$,
i.e. $p(l)=1/(N-1)$ recovers one of the basic constructions of the
SWN, by which connections to  all sites but the source are
equiprobable. For an infinite network, on the other hand, care has to
be exercised; in order to keep \eq{powerlaw} normalizable one has to
have $\alpha>1$.

We now turn to the basic procedure, in which the structure of the
model enters through its connectivity matrix ${\mathbb A}$. Now
${\mathbb A}$ is defined as follows: The off-diagonal  elements of
${\mathbb A}$, namely $A_{ij}$ with $i\neq j$, equal {\em minus} the
number of links between the sites $i$ and $j$ of the network. The
diagonal elements $A_{ii}$ count the total number of bonds connected
to $i$. Hence the connectivity matrix is symmetric and one has $\sum_i
A_{ij}=0$. Furthermore $\det({\mathbb A})=0$ and exactly one
eigenvalue of ${\mathbb A}$, say $E_1$, vanishes. One should note that
from the spectrum of the ${\mathbb A}$  matrix one can determine
e.g. the diffusion and vibrational properties of the structure, as
well as its behavior in external fields\cite{parbati,sj2}. The
spectrum of ${\mathbb A}$ for the SWN ($\alpha=0$) has been recently
studied by  Monasson\cite{monasson}; among his findings was the
existence of a pseudo-gap in the density of states, a property which
affects the long time diffusion properties\cite{sj2}. We note that
${\mathbb A}$ can be viewed as arising from two sources: One term,
${\mathbb A}^{(1)}$, is deterministic and is due to the underlying
regular lattice (here the ring). Another one, ${\mathbb A}^{(2)}$, is
stochastic and arises from the randomly added links. Thus ${\mathbb
A}={\mathbb A}^{(1)}+{\mathbb A}^{(2)}$. Now formally $A^{(1)}_{ii\pm
1}=-1$, $A^{(1)}_{ii}=2$ and $A^{(1)}_{ij}=0$ otherwise, where we
identify site $N+1$ with $1$ (periodic boundary conditions).  The
entries in $A^{(2)}_{ij}$ are random, and for $i\neq j$ equal $0$,
$-1$ or $-2$. In fact, letting $l$ be the chemical distance between
$i$ and $j$ ($i\neq j$), one has for the probability $P_l(c)$ that
$A^{(2)}_{ij}=-c$ 
\begin{equation}
\label{distr}
P_l(c)=\binom{2}{c}
\left(qal^{-\alpha}\right)^c\left(1-qal^{-\alpha}\right)^{2-c}
\end{equation}
The diagonal elements are, as before, determined from the requirement
that$\sum_i A^{(2)}_{ij}=0$. We close on a small note of caution by
remarking that, due to our prescription, even for $j\in
\{1,\ldots,i-1\}$, the elements $A^{(2)}_{ij}$  are not independent of
each other. Thus if $A^{(2)}_{ij}=-2$ for $i\neq j$, then for
$k\notin\{i,j\}$ one cannot have $A^{(2)}_{ik}=-2$; a nondiagonal
element having a value of $-2$ implies namely both for $j$ and for
$k$, that one of their additional bonds has started at $i$. By
construction, however, $i$ can only be the source of one additional
bond. For decreasing $q$ and increasing $N$, however, we expect such
correlations between the $A^{(2)}_{ij}$ to be less and less important. 

\section{Probability of Being at the Origin}\label{rel}
As a simple dynamical problem on the underlying structure, we focus
now on the probability for a random walker to be (still or again) at
the origin of its 
walk at a later time. This quantity is fundamental for fractal
lattices, where it leads directly to the spectral (harmonic)
dimension\cite{alexander}, a quantity of much
importance\cite{blumen}. As we have shown in a previous work, 
determining this quantity through a numerical cellular automaton
procedure is quite straightforward and very revealing for
SWN\cite{sj1}.  We look at the probability $P(i,t|m)$ for the walker
to be at site $i$ at time $t$, given that it started at site $m$ at
time $t=0$. One notes first that $P(i,t|m)$  obeys the following
master equation: 
\begin{equation}
\label{master}
\frac{\partial P(i,t|m)}{\partial t}=-\sigma\sum_{j=1}^{N} A_{ij}
P(j,t|m),
\end{equation}
where $\sigma$ is a transition rate. In vector notation
$\mathbf{P}^{(m)}(t)\equiv  (P(1,t|m),\ldots,P(N,t|m))$ this relation
reads:
\begin{equation}
\label{vmaster}
\frac{\partial {\mathbf P}^{(m)}(t)}{\partial t}=-\sigma {\mathbb A}
{\mathbf P}^{(m)}(t).
\end{equation}
and has the formal solution:
\begin{equation}
\label{sol1}
{\mathbf P}^{(m)}(t)=\exp\left(-\sigma {\mathbb A}t\right){\mathbf
P}^{(m)}(0).
\end{equation}
Now the initial condition is $\mathbf
P^{(m)}(0)=(0,\ldots,1,\ldots,0)$ with a single non-zero element at
$m$. The probability that the walker is again at $m$ at time $t$ reads
\begin{equation}
\label{sol2}
P(m,t|m)=[{\mathbf P}^{(m)}(t)]_m=\sum_j \left[\exp\left(-\sigma
{\mathbb A}t\right)\right]_{mj}[{\mathbf
P}^{(m)}(0)]_j=\left[\exp\left(-\sigma {\mathbb A}t\right)\right]_{mm} 
\end{equation}
This expression simplifies by averaging over all starting points,
since then
\begin{equation}
\label{relax}
\frac{1}{N}\sum_{m=1}^{N}
P(m,t|m)=\frac{1}{N}\mbox{Tr}\left(\exp\left(-\sigma {\mathbb
A}t\right)\right)=\frac{1}{N}\sum_{i=1}^{N} e^{-E_i\sigma t}
\end{equation}
holds, where Tr denotes the trace operation and $E_i$ with $1\leq
i\leq N$ are the eigenvalues of the (symmetric) connectivity matrix
${\mathbb A}$.  Note that in \eq{relax} because of the averaging over
all initial points only the eigenvalues enter. Furthermore one can now
readily average over different realizations, obtaining (since
$E_1\equiv 0$):
\begin{equation}
\label{avrelax}
P_0(t)\equiv\frac{1}{N}+\frac{1}{N}\langle\sum_{i=2}^{N} e^{-E_i\sigma
t}\rangle=\frac{1}{N}+\int\rho(E)e^{-Et}dE
\end{equation}
with $\rho(E)$ being the spectral density for $E>0$. 

\begin{figure}[h]
\unitlength=1cm
\begin{center}
\begin{picture}(6,5.8)
\put(-2.2,6.6){ \includegraphics{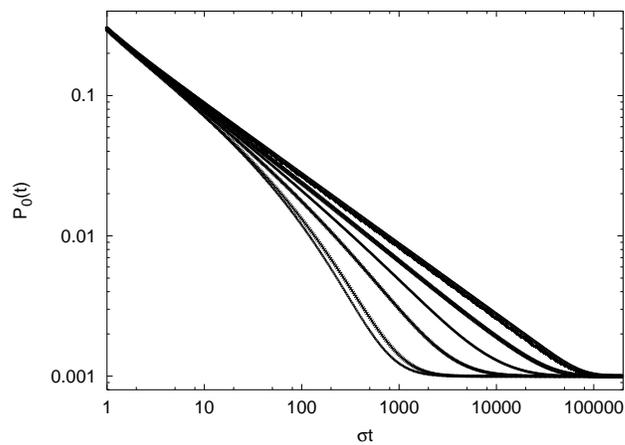}}
\end{picture}
\end{center}
\caption{The probability of being at the origin,
  $P_0(t)$, for $q=0.05$ and $\alpha=3,2,1.5,1.25,1,0.5$ and $0.0$ from
  upper to lower right. The curves for $\alpha=2$ and $\alpha=3$ are
  hardly distinguishable in the figure.}  
\label{p1}
\end{figure}

We turn now to our calculations, by which we determine numerically
$P_0(t)$ for different choices of $\alpha$ and $q$. For systems of
size $N=1001$ we construct the connectivity matrix, diagonalize it,
and employ Eq. (\ref{relax}) to evaluate $P_0(t)$. We use for each
choice of $\alpha$ and $q$ $100$ realizations to average over the
structural disorder. In Fig. (\ref{p1}) we display on double
logarithmic  scales $P_0(t)$ versus the dimensionless time $\sigma t$
for $q=0.05$ and for $\alpha$ ranging from $0$ to $3$. 
First we note that for very long times, $P_0(t)$ reaches the constant
value $1/N$, which arises due to the eigenvalue $E_1=0$. Increasing
the size of the {\swn} (i.e. $N$) pushes the long time 
plateau to lower values, but, as we have shown in an earlier work for
$\alpha=0$\cite{sj1}, leaves the $P_0(t)$ curves above the plateau
practically unaffected. This is also what we find here for general
$\alpha$; this allows us to infer the qualitative features of $P_0(t)$
in the limit $N\rightarrow\infty$. 

Turning now to the discussion of the results, we note first that for
$\alpha=0$ they agree perfectly with our previous {\swn}
analysis\cite{sj1}, which was based not on the diagonalization of
$\mathbb A$ but on a cellular automaton method. The decay of $P_0(t)$
follows at early times a power-law, which turns at later times into a
stretched exponential behavior. Asymptotically, the decay obeys to
leading order the form $\exp(-Ct^{1/3})$\cite{sj1,sj2} (with $C$ a
constant), which follows from the spectral density of
Ref. \cite{monasson}, $\rho(E)\sim E^{-1/2}\exp(-C'E^{-1/2})$.  Note
that in Fig.~\ref{p1} the curves flatten with increasing $\alpha$, a
sign that with growing $\alpha$, a walker is less prone to get far
away from its starting site. At early times $P_0(t)$ is little
affected by variations in $\alpha$, since at very short times it does
not matter whether the AL bring the walkers very far away or not. The
transition to pure power-law behavior appears to happen roughly around
$\alpha=2$. 

\begin{figure}[h]
\unitlength=1cm
\begin{center}
\begin{picture}(6,5.8)
\put(-2.2,6.6){ \includegraphics{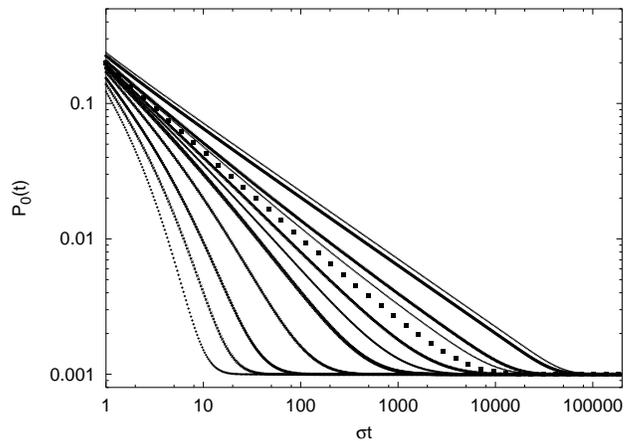}}
\end{picture}
\end{center}
\caption{Same as Fig. \ref{p1}, but for a choice of $q=0.8$. From
  upper to lower right are
  $\alpha=5.0,3.0,2.2,2.1,2.0,1.9,1.8,1.7,1.5,1.0,0.5$ and $0.0$; the
  curve for $\alpha=2$ is indicated by dots.}
\label{p2}
\end{figure}

Moving on to larger $q$ to examine whether this transition depends on
$q$, we plot in Fig. \ref{p2} the results for $q=0.8$.  
In this case there are more AL, and the results are more
sensitive to the value of $\alpha$: in Fig.~\ref{p2} the cases
$\alpha=3.0$ and $\alpha=5.0$ are easily distinguished. However, increasing
$\alpha$ further does not change the curves significantly.
The curves are also more spread out in the short-time domain than in
the case $q=0.05$. This is due to the fact that the quasi
1-dimensional behavior of the walk is mainly felt on distances of the
order of $1/q$, this being a measure of the mean distance between
branching points\cite{sj1}. We hasten to note that for very
small times (not displayed on the Fig.~\ref{p2}) the curves for different
$\alpha$ do coincide.
Despite these differences, the qualitative behavior of the curves on
Fig.~\ref{p1} and Fig.~\ref{p2} is comparable. Furthermore, the
cross-over behavior of 
$\alpha=2$ (given as a dotted line) appears even more clearly on Fig.~\ref{p2}: the curves with $\alpha>2$ follow power-law decays closely,
while the curves for $\alpha<2$ are partly concave, thus displaying a
faster than power-law decay.

\section{Typology of Random Walks}\label{discuss}
Let us briefly recall some terminology from the theory of random
walks. A random walk is said to be recurrent, if it returns with
probability $1$ to the origin at some later time. Otherwise the walk
is called transient. For a walk to be transient requires an infinite
system, because in finite systems all walks (disregarding situations
with traps, mortal walkers, etc) are recurrent. On homogenous
lattices, a walk is transient if and only if
\begin{equation}
\label{transient}
I\equiv\int_0^\infty dt\,P_0(t)
\end{equation}
is finite. In line with this, we could expect walks on the GSWN with
small $\alpha$ to be transient, given that for a stretched exponential behavior ($\beta>0$):
\begin{equation}
\label{swn}
I\sim\int^\infty dt\, \exp\left(-C t^{\beta}\right)<\infty.
\end{equation}
Moreover, in the opposite limit of large $\alpha$, we observe that
$P_0(t)$ follows a power-law decay with the exponent being nearly
$-1/2$. It follows that for large $\alpha$ we have $I=\infty$, an indication
that the walk is recurrent.  As we discuss in detail in the following,
for walks on an 
infinite regular linear chain whose steps are long-ranged and obey
\eq{powerlaw} for the step-lengths, the transition between recurrence
and transience occurs at $\alpha=2$\cite{hughes}.  It is now tempting
to aim at explaining our findings of Sec.~\ref{rel} along such
lines. Such a connection is achieved by replacing the random
{\swn}-structure under 
investigation here by a regular one (a mean-field type approach), and
letting the transition rates reflect the underlying statistics of the
links\cite{pandit2}. In this way the probability of taking a step of
length $l>1$ is proportional to $l^{-\alpha}$. 
However, as we show in the following, this regularizing approach is
not particularly succesful, since it does not describe $P_0(t)$ well
for small and moderate $\alpha$. 

We start now from the so-called the Riemann walks\cite{hughes}, which
are symmetric random walks on the linear chain, where each step of the
walk can extend over the length $l$ with probability
\begin{equation}
\label{walk}
\pi(l)\sim l^{-\alpha},\hspace{1cm}\alpha>1
\end{equation}
Such walks are recurrent for $\alpha\geq 2$ and transient for
$\alpha<2$\cite{hughes}. Riemann-walks are examples from the more
general class of L\'evy Flights and L\'evy Walks\cite{klafter}.
Turning now to the problem of averaging both the GSWN structures and
the random-walks over them, we simply replace in \eq{master} $\mathbb
A^{(2)}$ (remember that ${\mathbb A}= {\mathbb A}^{(1)}+{\mathbb
  A}^{(2)}$) by its average $\langle\mathbb A^{(2)}\rangle$ over all
GSWN. For the averaging we may use \eq{distr} and obtain
\begin{equation}
\label{meanfield}
\langle A^{(2)}_{ij}\rangle=2al^{-\alpha}q\equiv c(l),
\end{equation}
where $l$ is the chemical distance between $i$ and $j$. By doing this we
have now as connectivity matrix ${\mathbb A}={\mathbb
A}^{(1)}+\langle{\mathbb A}^{(2)}\rangle$, whose eigenvalues $E_k$
are readily found; they read (for $N$ odd):
\begin{equation}
\label{eigenvalues}
E_k=2-2\cos(2\pi k/N)+2\sum_{j=1}^{(N-1)/2} (1-\cos(2\pi kj/N)) c(j),
\end{equation}
where $k=0\ldots N-1$. Now \eq{eigenvalues} can be used to determine
numerically $P_0(t)$ via Eq. (\ref{avrelax}). 

\begin{figure}[h]
\unitlength=1cm
\begin{center}
\begin{picture}(6,5.8)
\put(-2.2,6.6){ \includegraphics{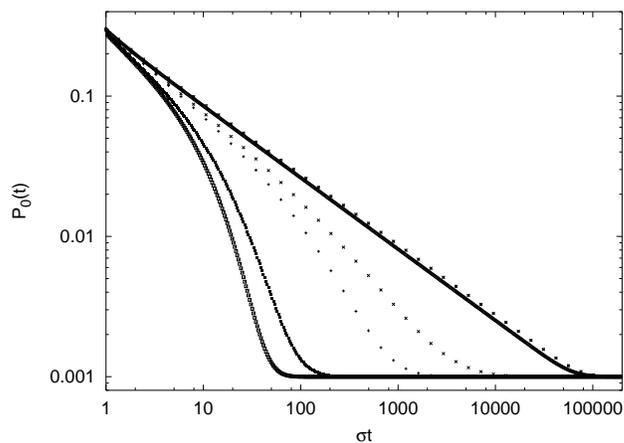}}
\end{picture}
\end{center}
\caption{Comparison of the mean field theory prediction (full curves) with the
  numerical data (dotted curves) for $q=0.05$. The values of $\alpha$
  are from above $\alpha=3$, $\alpha=1.5$ and $\alpha=0$. Good
  agreement is found only for large $\alpha$, here $\alpha=3$.}
\label{mft}
\end{figure}

In Fig.~\ref{mft} we compare for $N=1001$, $q=0.05$ and $\alpha=3$,
$\alpha=1.5$ and $\alpha=0$ the results of the two approaches. 
For $\alpha=3$ the two methods lead to a nice agreement; it seems
that for $\alpha$ around or larger than $3$ the 
fluctuations due to the disorder play only a minor role.
On the other hand, as exemplified by $\alpha=1.5$ and $\alpha=0$,
for $\alpha$ below $2$ the mean field approach leads to
$P_0(t)$-forms which are quite different from those obtained in
Sec.~\ref{rel}.

\section{Conclusion}\label{conclusion}
In this work we have studied a new variant of the {\swn} (SWN) model
which takes into account the fact that the probability of adding links
can depend on the chemical distance between the connected
sites. Exemplarily, here we have taken the probability
distribution to be a power law (with exponent $\alpha$) of the chemical
distance, see \eq{powerlaw}. We have focused on random walks
and especially on the probability $P_0(t)$ of a random walker to be at
its origin. Depending on the value of $\alpha$ we have found
qualitatively different behaviors. Specifically, we found
clues indicating
that in the infinite system limit random walks on GSWN may change from being
transient to being recurrent, as $\alpha$ crosses the marginal
value of $2$ from below. Moreover, we have shown that our
model is related to L\'evy flights and to Riemann walks. We also
found that a simple mean-field regularization of the GSWN-problem
gives poor results for small $\alpha$. Overall, it follows that GSWN with
$\alpha<2$ are objects whose dynamical properties differ significantly
from those of regular lattices.

\renewcommand{\abstractname}{Acknowledgements}
\begin{abstract}
The support of the DFG, of the GIF through grant I0423-061.14, and of
  the Fonds der Chemischen Industrie are gratefully acknowledged.
\end{abstract}

\end{document}